\pdfoutput=1
\RequirePackage{ifpdf}
\ifpdf % We are running pdfTeX in pdf mode
\documentclass[pdftex]{sigma}
\else
\documentclass{sigma}
\fi

\usepackage{accents}
\numberwithin{equation}{section}

\begin{document}

\def\bra#1{\left\langle #1\right|}
\def\ket#1{\left| #1\right\rangle}
\def\bracket#1#2{\left\langle #1  \,\right| \left. #2 \right\rangle}
\def\FL#1{\left\lfloor #1 \right\rfloor}

\allowdisplaybreaks

\newcommand{\arXivNumber}{1408.4842}

\renewcommand{\PaperNumber}{002}

\FirstPageHeading

\ShortArticleName{Lowest Weight Representations of Conformal Galilei Algebras}

\ArticleName{Lowest Weight Representations, Singular Vectors\\
and Invariant Equations for a~Class of Conformal\\
Galilei Algebras}

\Author{Naruhiko AIZAWA~$^\dag$, Radhakrishnan CHANDRASHEKAR~$^\ddag$ and Jambulingam SEGAR~$^\S$}

\AuthorNameForHeading{N.~Aizawa, R.~Chandrashekar and J.~Segar}

\Address{$^\dag$~Department of Mathematics and Information Sciences, Osaka Prefecture University,\\
\hphantom{$^\dag$}~Nakamozu Campus, Sakai, Osaka 599-8531, Japan}
\EmailD{\href{mailto:aizawa@mi.s.osakafu-u.ac.jp}{aizawa@mi.s.osakafu-u.ac.jp}}

\Address{$^\ddag$~Department of Physics, National Chung Hsing University, Taichung 40227, Taiwan}
\EmailD{\href{mailto:chandrashekar10@gmail.com}{chandrashekar10@gmail.com}}

\Address{$^\S$~Department of Physics, Ramakrishna Mission Vivekananda College,\\
\hphantom{$^\S$}~Mylapore, Chennai 600 004, India}
\EmailD{\href{mailto:segar@imsc.res.in}{segar@imsc.res.in}}

\ArticleDates{Received August 22, 2014, in f\/inal form December 31, 2014; Published online January 06, 2015}

\Abstract{The conformal Galilei algebra (CGA) is a~non-semisimple Lie algebra labelled by two parameters $d$ and $\ell$.
The aim of the present work is to investigate the lowest weight representations of CGA with $d = 1$ for any integer
value of $ \ell$.
First we focus on the reduci\-bi\-li\-ty of the Verma modules.
We give a~formula for the Shapovalov determinant and it follows that the Verma module is irreducible if $\ell = 1$ and
the lowest weight is nonvanishing.
We prove that the Verma modules contain many singular vectors, i.e., they are reducible when $\ell \neq 1$.
Using the singular vectors, hierarchies of partial dif\/ferential equations def\/ined on the group manifold are derived.
The dif\/ferential equations are invariant under the kinematical transformation generated by CGA.
Finally we construct irreducible lowest weight modules obtained from the reducible Verma modules.}


\Classification{17B10; 58J70}

\section{Introduction}

Conformal algebras are algebraic structure relevant to physical problems for both relativistic and non-relativistic
settings.
In the non-relativistic setting, the algebra is called conformal Galilei algebra (CGA)~\cite{NdORM1} (see
also~\cite{HavPre}).
This is a~family of Lie algebras consisting of inf\/initely many members.
Each member is not semisimple, not isomorphic to each other and is labelled by two parameters $d$ and $\ell$,
where~$d$ is a~positive integer and $ \ell $ takes a~spin value $(=1/2, 1, 3/2, \dots)$.
The simplest member with $\ell=1/2$ is the Schr\"odinger algebra which gives the symmetry algebra of free Schr\"odinger
equations in~$d$-dimensional space~\cite{Hagen, Nie}.
Recently the Schr\"odinger algebra and $ \ell=1$ member of CGA were discussed in the context of non-relativistic AdS/CFT
correspondence~\cite{AlDaVa,BaGop,BaMcG,MT, Son}.
This caused a~renewed interest on CGA.
Indeed, CGA with various pairs of $(d,\ell)$ appears in wide range of physical
problems~\cite{AndGaGoMa,AndGo,AndGo2,AndGoKD,AndGoKoMa,AndGoMas,GalaMas,GalaMas2,GalaMas3,Henkel,HuJa,LSZ,NisSon,StZak,ZhangHor}
(see~\cite{AiIsKi} for more references on $\ell=1/2 $ and $ \ell =1 $ CGA).
This observation motivates us to study representation theory of CGA.

This work is a~continuation of our previous works on the lowest (or highest) weight representations of CGA and their
application to the symmetry of partial dif\/ferential equations~\cite{AiIsKi,AiKimSeg}.
In~\cite{AiIsKi,AiKimSeg} we focused on $ d=1$ CGA with half-integer~$\ell$.
On the other hand, we put our focus on $d = 1$ CGA with \textit{integer}~$\ell$ in this work.
Our aim is two fold: (i)~we study some important ingredients of representation theory such as Verma modules, singular
vectors and so on.
We start with the Verma modules and study their reducibility by searching for the singular vectors.
If the Verma module is reducible, we take a~quotient by the maximal invariant submodule and study the reducibility of
the quotient module by the same method.
This process is repeated until we reach the irreducible module.
At each step of the process we would like to give the explicit formula for the basis of the modules.
The formula is independent of any particular realization of the generators of the algebra.
This is a~nice and useful feature when we consider the application of a~representation theory, since we can take
a~realization appropriate to the problem under consideration.
Indeed, our second aim shall be achieved by taking a~vector f\/ield realization on a~group manifold.
(ii)~We construct a~hierarchy of partial dif\/ferential equations by employing the method developed in~\cite{Dobrev}.
By construction, these equations have kinematical symmetries generated by the CGA.
This means that the CGA generates a~transformation of independent variables (dependent variable remains unchanged)
through which a~solution to the equation is converted to another solution.
For any integer values of~$\ell$ each hierarchy contains one dif\/ferential equation of second order.
This is in sharp contrast to the previous works on dynamical systems relating to the CGA, where higher order derivative
is needed for higher values of~$\ell$~\cite{AndGaGoMa,AndGo,AndGoKD,AndGoMas}.

Here we mention the previous works on irreducible lowest/highest weight modules and dif\/fe\-ren\-tial equations with
symmetries generated by CGA for other pairs of $(d,\ell)$.
A~classif\/ication of irreducible modules for the $ \ell = 1/2 $ and $ d = 1, 2, 3 $ is found in~\cite{DoDoMr, Mrugalla},
while the case $ (d,\ell) = (2,1) $ is found in~\cite{AiIs}.
Symmetric dif\/ferential equations have been obtained for $ \ell=1/2 $ and any~$d$~\cite{ADD,ADDS,DoDoMr, Mrugalla}.
Such equations were also found for $ d = 1, 2 $ and any half-integer~$\ell$~\cite{AiKimSeg}.
Very recently a~classif\/ication of all f\/inite weight modules over the $ d=1 $ CGA with any $ \ell $ has been done
in~\cite{LuMazoZhao}.

This paper is organized as follows.
In the next section the def\/inition of $d = 1$ CGA for integer~$\ell$ is given.
We also give triangular type decomposition and algebraic anti-involution for later use.
In Section~\ref{SEC:VM} the Verma modules over the CGA is introduced and calculation of the Shapovalov determinant is
presented.
We observe that the Shapovalov determinant vanishes for many cases.
This suggests the existence of singular vectors in the Verma modules.
It is shown in Section~\ref{Sec:SV} that there exist various singular vectors.
It turns out that there exists more singular vectors for integer~$\ell$ than for half-integer~$\ell$.
The formulae of singular vectors are used to construct the hierarchy of partial dif\/ferential equations with kinematical
symmetries in Section~\ref{SED:EQS}.
In Section~\ref{SEC:IRREP} we give the irreducible lowest weight modules obtained from the Verma module.
Throughout this article we denote the $ d= 1 $ CGA with spin~$\ell$ by~$ {\mathfrak g}_{\ell}$.

\section[$d=1$ Conformal Galilei algebras ${\mathfrak g}_{\ell}$]{$\boldsymbol{d=1}$ Conformal Galilei algebras $\boldsymbol{{\mathfrak g}_{\ell}}$}

The complex Lie algebra $ {\mathfrak g}_{\ell} $ for a~f\/ixed integer $ \ell $ has the elements~\cite{NdORM1}:
\begin{gather*}
D, \ H, \ C, \ P_n, \quad n = 0, 1,   \dots,   2\ell.
%\label{generators}
\end{gather*}
Their nonvanishing commutators are given~by
\begin{alignat}{4}
& [D, H] = H, \qquad & & [D, C] = - C, \qquad & & [C, H] = 2D,& \nonumber\\
& [H, P_n] = -n P_{n-1}, \qquad & & [D, P_n] = (\ell-n) P_n, \qquad & & [C, P_n] = (2\ell-n) P_{n+1}.&\label{nonvanishing-commutators}
\end{alignat}
One may see from this that $ \langle   P_0, P_1, \dots, P_{2\ell}   \rangle $ is an Abelian ideal of $ {\mathfrak
g}_{\ell} $, so that $ {\mathfrak g}_{\ell} $ is \textit{not} semisimple.
It is known that this Lie algebra has no central extensions~\cite{MT}.
The subalgebra spanned by $ \langle
 H, D, C
\rangle $ is isomorphic to $ {\mathfrak{so}}(2,1) \simeq {\mathfrak{sl}}(2,{\mathbb R}) \simeq {\mathfrak{su}}(1,1)$.
The Abelian subalgebra spanned by $ \langle
P_n
\rangle_{n=0, 1, \dots, 2\ell} $ carries the spin $ \ell $ representation of the $ {\mathfrak{sl}}(2,{\mathbb R})$
subalgebra.

This algebra may be realized as generators of transformation of $(1+1)$-dimensional spacetime
\begin{gather*}
H = \frac{\partial}{\partial t},
\qquad
D =-t\frac{\partial}{\partial t} -\ell x \frac{\partial}{\partial x},
\qquad
C = t^2 \frac{\partial}{\partial t} + 2 \ell tx \frac{\partial}{\partial x},
\qquad
P_n = (-t)^n \frac{\partial}{\partial x}.
%\label{CGArealization-in-space}
\end{gather*}
In this realization,~$H$,~$D$ and~$C$ generates time translation, dilatation and the special conformal representation,
respectively.
Meanwhile the generator of space translation is represented by~$P_0$, the Galilei transformation generator by~$P_1$,
the transformation to a~reference frame with constant acceleration is given by~$P_2$ and so on.

One may introduce the algebraic anti-involution $ \omega: {\mathfrak g}_{\ell} \to {\mathfrak g}_{\ell} $~by
\begin{gather*}
\omega(D) = D, \qquad \omega(H) = C, \qquad \omega(P_n) = P_{2\ell-n}.
%\label{anti-inv}
\end{gather*}
It is not dif\/f\/icult to verify that $ \omega $ satisf\/ies the required relations:
\begin{gather*}
\omega([X,Y]) = [\omega(Y), \omega(X)], \qquad \omega^2(X) = X \qquad  \forall\, X \in {\mathfrak g}_{\ell}.
%\label{anti-inv2}
\end{gather*}
Let us def\/ine the degree of the generators based on their commutator with respect to~$D$:
\begin{gather*}
\deg(D) = 0, \qquad \deg(H) = 1, \qquad \deg(C) = -1, \qquad \deg(P_n) = \ell-n.
%\label{degree}
\end{gather*}
With respect to the sign of the degree one may def\/ine the triangular decomposition of ${\mathfrak g}_{\ell}$:
\begin{gather*}
    {\mathfrak g}_{\ell} = {\mathfrak g}_{\ell}^+ \oplus {\mathfrak g}_{\ell}^0 \oplus {\mathfrak g}_{\ell}^-,
%\label{tri-decomp}
\end{gather*}
where
\begin{gather*}
{\mathfrak g}_{\ell}^+ = \langle H, P_0, P_1, \dots, P_{\ell-1} \rangle,
\qquad
{\mathfrak g}_{\ell}^0 = \langle D, P_{\ell} \rangle,
\qquad
{\mathfrak g}_{\ell}^- = \langle C, P_{\ell+1}, P_{\ell+2}, \dots, P_{2\ell} \rangle.
\end{gather*}
This is a~decomposition of $ {\mathfrak g}_{\ell} $ as a~direct sum of the vector spaces.

\section{Verma modules and Shapovalov determinant}
\label{SEC:VM}

We start with the one-dimensional module $ {\mathbb C} \ket{\delta,p} $ over $ {\mathfrak b} = {\mathfrak g}_{\ell}^0
\oplus {\mathfrak g}_{\ell}^-$. The lowest weight vector $ \ket{\delta,p} $ is def\/ined~by
\begin{gather*}
D \ket{\delta,p} = \delta \ket{\delta,p},
\qquad
P_{\ell} \ket{\delta,p} = p \ket{\delta,p},
\qquad
X \ket{\delta,p} = 0,
\qquad
 \forall \, X \in {\mathfrak g}_{\ell}^-.
%\label{LWV}
\end{gather*}
Then for each pair of $(\delta,p) $ the Verma module over $ {\mathfrak g}_{\ell} $ is def\/ined as usual (see, e.g.,~\cite{Dix}): $ V^{\delta,p}_{\ell} = U({\mathfrak g}_{\ell}) \otimes_{U({\mathfrak b})} \ket{\delta,p} $ where $U({\mathfrak g}_{\ell})$ and $ U({\mathfrak b}) $ are the enveloping algebras of $ {\mathfrak g}_{\ell} $ and $
\mathfrak b$, respectively.
In order to specify the basis of $ V^{\delta,p}_{\ell} $ we introduce the $ \ell$-component vector
\begin{gather*}
\underaccent{\tilde}{m} = (m_1, m_2, \dots, m_{\ell})  \in {\mathbb R}^{\ell},
%\label{m-vec-def}
\end{gather*}
where $ m_i$ $(i=1,2,\dots,\ell) $ are non-negative integers.
With this the basis of $ V^{\delta,p}_{\ell} $ is given~by
\begin{gather}
\ket{k,\underaccent{\tilde}{m}} = H^k P_{\ell-1}^{m_1} P_{\ell-2}^{m_2} \cdots P_0^{m_{\ell}} \ket{\delta,p}.
\label{anyl-basis}
\end{gather}
We also introduce the~$\ell$-component vectors
\begin{gather*}
\underaccent{\tilde}{\epsilon}_j = (0, \dots, 0, 1, 0, \dots,0) \in {\mathbb R}^{\ell},
\qquad
j=1,2,\dots, \ell.
%\label{tilde-epsilon-def}
\end{gather*}
where the~$j$th entry of $\underaccent{\tilde}{\epsilon}_j $ is~1, and all other entries are~0.
It may not be dif\/f\/icult to prove the following relation by induction on $k$:
\begin{gather}
[P_n, H^k] = \sum\limits_{i=1}^{\min\{n,k\}} \binom{n}{i} \frac{k!}{(k-i)!}  H^{k-i}   P_{n-i},
\qquad
n \geq 1.
\label{PnHkcomm}
\end{gather}
It follows that the action of $ {\mathfrak g}_{\ell}^0$ and ${\mathfrak g}_{\ell}^+$ on $\ket{k,\underaccent{\tilde}{m}}$:
\begin{gather*}
    D \ket{k,\underaccent{\tilde}{m}} = \left(\delta + k+ \sum\limits_{i=1}^{\ell} i   m_i\right) \ket{k,\underaccent{\tilde}{m}},
\nonumber
\\
    P_{\ell} \ket{k,\underaccent{\tilde}{m}} = p \ket{k,\underaccent{\tilde}{m}} + \sum\limits_{i=1}^{\min\{\ell,k\}}
\binom{\ell}{i} \frac{k!}{(k-i)!} \ket{k-i,\underaccent{\tilde}{m}+\underaccent{\tilde}{\epsilon}_{i}},
\nonumber
\\
    H \ket{k,\underaccent{\tilde}{m}}= \ket{k+1,\underaccent{\tilde}{m}},
\nonumber
\\
    P_n \ket{k,\underaccent{\tilde}{m}} = \sum\limits_{i=1}^{\min\{n,k\}} \binom{n}{i} \frac{k!}{(k-i)!}
\ket{k-i,\underaccent{\tilde}{m}+\underaccent{\tilde}{\epsilon}_{\ell-n+i}},
\qquad
1 \leq n \leq \ell-1,
\nonumber
\\
    P_0 \ket{k,\underaccent{\tilde}{m}} = \ket{k,\underaccent{\tilde}{m}+\underaccent{\tilde}{\epsilon}_{\ell}}.
%\label{g0p-action-any}
\end{gather*}
From this one see that $ V^{\delta,p}_{\ell} $ has a~grading structure according to the eigenvalue of~$D$:
\begin{gather*}
    V^{\delta,p}_{\ell} = \bigoplus_{N=0}^{\infty} \big(V^{\delta,p}_{\ell}\big)_N,
\qquad
\big(V^{\delta,p}_{\ell}\big)_N = \big\{ \ket{v} \in V^{\delta,p}_{\ell} \, | \, D \ket{v} = (\delta+N) \ket{v} \big\}.
%\label{Verma-grading}
\\
    N = k+ \sum\limits_{i=1}^{\ell} i   m_i.
%\label{N-definition}
\end{gather*}
We refer to~$N$ as the \textit{level} as usual.

Now we def\/ine Shapovalov form on a~module over $ {\mathfrak g}_{\ell} $~\cite{Shapo} (see also, e.g.,~\cite{KacRai}).
That is a~contravariant Hermitian form $ \bracket{\cdot}{\cdot} $ def\/ined by making use of the anti-involution~ $\omega$.
For the Verma module $ V^{\delta,p}_{\ell} $ it is def\/ined as follows:   Let $ \ket{x} $ and $ \ket{y} $ be any two
vectors in $ V^{\delta,p}_{\ell}$.
They may be written as
\begin{gather*}
\ket{x} = X \ket{\delta,p},
\qquad
\ket{y} = Y \ket{\delta,p},
\qquad
X, Y \in U({\mathfrak g}_{\ell}^+).
\end{gather*}
Then
\begin{gather}
\bracket{x}{y} = \bra{\delta,p} \omega(X) Y \ket{\delta,p},
\qquad
\bracket{\delta,p}{\delta,p} = 1.
\label{innerprod-def}
\end{gather}
Next we def\/ine the Shapovalov determinant at level~$N$.
Let $ \ket{v_1}, \ket{v_2}, \dots, \ket{v_r} $ be a~set of basis of the subspace $ \big(V^{\delta,p}_{\ell}\big)_N$.
We consider the matrix $ (\bracket{v_i}{v_j}) $ whose entries are     the Shapovalov form and we call the determinant of
this matrix, the Shapovalov determinant at level $N: \Delta^{(\ell)}_N = \det (\bracket{v_i}{v_j})$.
We give an explicit formulae of the Shapovalov determinants of $ {\mathfrak g}_{\ell}$, since $ \Delta^{(\ell)}_N $ is
important to know the reducibility of $V^{\delta,p}_{\ell}$.
\begin{proposition}\label{prop1}
The Shapovalov determinants $ \Delta^{(\ell)}_N$ at level~$N$ of ${\mathfrak g}_{\ell}$ are given as follows $($up to
overall sign$)$:
\begin{alignat}{3}
   & i)
\quad &&
\Delta_N^{(1)} = \left(\prod\limits_{m=0}^N m! \right)^2 (2p)^{N(N+1)},&
\label{KDl1}
\\
  &   ii)
\quad &&
\Delta^{(\ell)}_N = (\ell+1)^2   p^2   \delta_{N1},
\qquad
\ell \geq 2.&
%\label{KDell}
\end{alignat}
\end{proposition}

\begin{proof}
The case with $ \ell = 1 $ shows a~deviation from other values of $ \ell$.
We treat the case with $ \ell = 1 $ separately.
The basis of $ V^{\delta,p}_1 $ is specif\/ied by two nonnegative integers:
\begin{gather*}
\ket{k,m} = H^k P_0^m \ket{\delta,p}.
\end{gather*}
The next relation obtained from~\eqref{PnHkcomm} is useful in the following computation:
\begin{gather}
P_2 \ket{k,m} = 2kp \ket{k-1,m} + k(k-1) \ket{k-2,m+1}.
\label{ActionP2}
\end{gather}
The level is $ N = k +m $ so that the basis of $ \big(V^{\delta,p}_1\big)_N $ is given~by
\begin{gather*}
\ket{m,N-m},
\qquad
0 \leq m \leq N.
\end{gather*}
The product of two vectors in $ \big(V^{\delta,p}_1\big)_N $ is
\begin{gather}
\bracket{N-k,k}{m,N-m} = \bra{N-k,0} P_0^{N-m} P_2^k \ket{m,0}.
\label{IPl=1}
\end{gather}

It follows from~\eqref{ActionP2} that $P_2^k \ket{m,0}$ is a~linear combination of $ \ket{m-k-j,j} $ with $0 \leq j
\leq k$.
Therefore, $ P_2^k \ket{m,0} = 0$ if $k > m$.
Thus we have proved the following lemma:

\begin{lemma}
\label{lemma1}
$ \bracket{N-k,k}{m,N-m} = 0 $ if $ k > m$.
\end{lemma}

By def\/inition $ \Delta^{(1)}_N $ is given by (up to sign)
\begin{gather*}
\Delta^{(1)}_N = \left|
\begin{matrix}
 \bracket{N,0}{0,N} & \bracket{N,0}{1,N-1} & \dots & \bracket{N,0}{N,0}
\\
\bracket{N-1,0}{0,N} & \bracket{N-1,0}{1,N-1} & \dots & \bracket{N-1,0}{N,0}
\\
\vdots & \vdots & \ddots & \vdots
\\
\bracket{0,N}{0,N} & \bracket{0,N}{1,N-1} & \dots & \bracket{0N}{N,0}
\end{matrix}
\right|.
\end{gather*}
By Lemma~\ref{lemma1}, $ \Delta^{(1)}_N $ is the determinant of upper triangular matrix.
Thus
\begin{gather*}
\Delta_N^{(1)} = \prod\limits_{m=0}^N \bracket{N-m,m}{m,N-m}.
\end{gather*}
Each factor is calculated by~\eqref{IPl=1} as follows
\begin{gather*}
\bracket{N-m,m}{m,N-m} = \bra{N-m,m} P_0^{N-m} P_2^m \ket{N-m,m} = (N-m)!   m!   (2p)^N,
\end{gather*}
where we used the relation obtained from~\eqref{ActionP2}: $ P_2^m \ket{m,0} = (2p)^m m ! \ket{0,0}$.
This completes the proof of~\eqref{KDl1}.

Now let us turn to the case with $\ell \geq 2$.
In this case $ \Delta^{(\ell)}_N \neq 0 $ only if $N = 1$.
As we shall see, this fact stems from that at least two rows of $ \Delta^{(\ell)}_N $ are proportional if $ N > 1$.
The basis of $(V^{\delta,p}_{\ell})_N $ is given by~\eqref{anyl-basis} with the constraint $  N = k+
\sum\limits_{i=1}^{\ell} i   m_i$.
First we prove the following lemma:
\begin{lemma}
\label{lemma2}
In the subspace $ (V^{\delta,p}_{\ell})_N $ we have the relation
\begin{gather*}
\bracket{0,\underaccent{\tilde}{n}}{k,\underaccent{\tilde}{m}} = f^{(\ell)}_{N,\underaccent{\tilde}{n}}(p) \delta_{kN},
%\label{IPlge2}
\end{gather*}
where $ f^{(\ell)}_{N,\underaccent{\tilde}{n}}(p) $ is a~function of~$p$ determined by $ \ell$, $N $ and $
\ket{0,\underaccent{\tilde}{n}}$.
\end{lemma}
\begin{proof}
By def\/inition
\begin{gather}
    \bracket{0,\underaccent{\tilde}{n}}{k,\underaccent{\tilde}{m}} = \bra{\delta,p} P_{2\ell}^{n_{\ell}}
P_{2\ell-1}^{n_{\ell-1}} \cdots P_{\ell+1}^{n_1} H^k P_{\ell-1}^{m_1} \cdots P_0^{m_{\ell}} \ket{\delta,p},
\label{IPlge2-2}
\\
    N = \sum\limits_{i=1}^{\ell} i n_i = k + \sum\limits_{i=1}^{\ell} i  m_i.
\nonumber
\end{gather}
Because of~\eqref{PnHkcomm} the right hand side of~\eqref{IPlge2-2} will be a~linear combination of the terms
\begin{gather*}
\bra{\delta,p} H^{k'} P_{\ell-1}^{m'_1} \cdots P_0^{m'_{\ell}} \ket{\delta,p},
\end{gather*}
with the condition $ m'_k \geq m_k $ for all~$k$.
However, such terms give nonvanishing contributions only if
\begin{gather*}
k' = m'_1 = m'_2 = \dots = m'_{\ell} = 0.
\end{gather*}
This implies $ m_k = 0 $ for all $k$, i.e., $ k = N$.
\end{proof}

It follows from Lemma~\ref{lemma2} that two rows in $ \Delta^{(\ell)}_N $ labelled by $ \bra{0,\underaccent{\tilde}{n}}
$ and $ \bra{0,\underaccent{\tilde}{n'}} $ are proportional.
If $ N \geq 2 $ then there exists at least one such pair of rows, but no such pair for $N = 1$.
Thus $ \Delta^{(\ell)}_N = 0 $ if $ N \geq 2$.
On the other hand the $ N = 1 $ subspace is two-dimensional with the basis $ \ket{1, \underaccent{\tilde}{0}}, \ket{0,
\underaccent{\tilde}{\epsilon}_1} $ where $\underaccent{\tilde}{0}$ denotes the zero vector in ${\mathbb R}^{\ell}$.
Thus $ \Delta^{(\ell)}_1 $ is computed as follows
\begin{gather*}
\Delta^{(\ell)}_1 = \left|
\begin{matrix} \bracket{1, \underaccent{\tilde}{0}}{1, \underaccent{\tilde}{0}} & \bracket{1, \underaccent{\tilde}{0}}{0,
\underaccent{\tilde}{\epsilon}_1}
\vspace{1mm}\\  \bracket{0, \underaccent{\tilde}{\epsilon}_1}{1, \underaccent{\tilde}{0}} & \bracket{0,
\underaccent{\tilde}{\epsilon}_1}{0, \underaccent{\tilde}{\epsilon}_1}
\end{matrix}
\right| = \left|
\begin{matrix} 2 \delta & (\ell+1)p
\vspace{1mm}\\  (\ell+1)p & 0
\end{matrix}
\right| = - (\ell+1)^2   p^2   .
\end{gather*}
This completes the proof of Proposition~\ref{prop1}.
\end{proof}

\begin{proposition}\qquad
\label{prop2}
\begin{enumerate}\itemsep=0pt
\item[$i)$] $ V^{\delta,p}_{\ell} $ is irreducible if $ \ell = 1 $ and $ p \neq 0$.
\item[$ii)$] $ V^{\delta,p}_{\ell} $ is reducible if $ \ell \geq 2 $ and $ p \neq 0$.
\item[$iii)$] $ V^{\delta,0}_{\ell} $ is reducible for all values of $ \ell$.
\end{enumerate}
\end{proposition}

\begin{proof}
(i) is the corollary of Proposition~\ref{prop1}(i).
Proposition~\ref{prop1} also suggests the existence of singular vectors in $V^{\delta,0}_{\ell} $ for any $\ell$ and
in $V^{\delta,p}_{\ell}$ with $\ell \geq 2$ and $p \neq 0$.
In the next section we shall show that this is indeed the case.
Thus we establish~(ii) and~(iii).
\end{proof}

\section[Singular vectors in $V^{\delta,p}_{\ell}$]{Singular vectors in $\boldsymbol{V^{\delta,p}_{\ell}}$}
\label{Sec:SV}

In this section we give explicit formulae of the singular vectors in $ V^{\delta,p}_{\ell}$.
We do not give a~complete list of singular vectors, but the list given below is enough to show the reducibility of $V^{\delta,p}_{\ell}$ (Proposition~\ref{prop2}) and derive dif\/ferential equations having ${\mathfrak g}_{\ell}$ as
a~symmetry in the next section.
Before giving the list let us recall the def\/inition of singular vectors.
A~singular vector $\ket{v_s} \in V^{\delta,p}_{\ell}$ is yet another lowest weight vector which is not proportional to
$\ket{\delta,p}$.
Namely, $ \ket{v_s} $ satisf\/ies the conditions:
\begin{gather}
    \ket{v_s} \neq {\mathbb C} \ket{\delta,p},
\qquad\!\!
    D \ket{v_s} = \delta' \ket{v_s},
\qquad\!\!
P_{\ell} \ket{v_s} = p' \ket{v_s},
\qquad\!\!
X \ket{v_s} = 0,
\qquad\!\!
\forall \,  X \in {\mathfrak g}_{\ell}^-.\!\!\!
\label{SV-def}
\end{gather}
According to Proposition~\ref{prop1}, singular vectors in $V^{\delta,p}_{\ell}$ may be dif\/ferent for $p \neq 0$ and
$p = 0$.
We treat these cases separately.

\subsection[$p \neq 0$]{$\boldsymbol{p \neq 0}$}

In this case the Verma modules over $ \ell = 1 $ algebra have no singular vectors.
For the algebra $ {\mathfrak g}_{\ell} $ with $ \ell \geq 2 $, singular vectors may exist only in the subspace $
\big(V^{\delta,p}_{\ell}\big)_N $ with $ N \geq 2$.

\begin{proposition}
\label{prop3}
Following are the singular vectors in $V^{\delta,p}_{\ell} $ for $ \ell \geq 2$
\begin{gather}
\big({\cal S}^{(2n)}\big)^k \ket{\delta,p},
\qquad
\big({\cal S}^{(2n+1)}\big)^k \ket{\delta,p},
\qquad
k = 1, 2, \dots,
\label{SVformula1}
\end{gather}
where $ n $ takes the value of a~positive integer.
The maximal value of~$n$ is determined by $ \ell $ and~$N$ in such a~way that $ {\cal S}^{(2n)}$ and $ {\cal
S}^{(2n+1)} $ given below are well-defined
\begin{gather}
    {\cal S}^{(2n)} = p P_{\ell-2n} + \sum\limits_{j=1}^{n-1} a_j P_{\ell-2n+j} P_{\ell-j} + \frac{(-1)^n}{2}
\frac{\ell!  (\ell+2n)!}{((\ell+n)!)^2}  P_{\ell-n}^2,
\label{SVformula2}\\
 {\cal S}^{(2n+1)} = p^2 P_{\ell-2n-1} - \frac{\ell+2n+1}{(n+\frac{1}{2}) (\ell+1)} P_{\ell-1} S^{(2n)}  \nonumber
\\
\hphantom{{\cal S}^{(2n+1)}=}{}
- \frac{n-\frac{1}{2}}{n+\frac{1}{2}}  \frac{\ell+2n+1}{\ell+1}  p  P_{\ell-1}  P_{\ell-2n} -
\sum\limits_{j=1}^{n-1} b_j  P_{\ell-2n+j}  P_{\ell-j-1},
\label{SVformula3}
\end{gather}
where
\begin{gather*}
a_j = (-1)^j \frac{\ell!  (\ell+2n)!}{(\ell+j)!  (\ell+2n-j)!},
\qquad
b_j = (-1)^j \frac{n-\frac{1}{2}-j}{n+\frac{1}{2}} \frac{p \ell!  (\ell+2n+1)!}{(\ell+j+1)! (\ell+2n-j)!}.
\end{gather*}
\end{proposition}

\begin{proof}
We show that the vectors in~\eqref{SVformula1} satisfy the conditions in~\eqref{SV-def}.
It is obvious that the vectors in~\eqref{SVformula1} are annihilated by $ P_a \in {\mathfrak g}_{\ell}^- $ and are
eigenvectors of $P_{\ell}$ with the eigenvalue~$p$.
It is also easy to verify that
\begin{gather*}
\big[D, \big({\cal S}^{(2n)}\big)^k \big] = 2nk \big({\cal S}^{(2n)}\big)^k,
\qquad
\big[D, \big({\cal S}^{(2n+1)}\big)^k \big] = (2n+1)k \big({\cal S}^{(2n+1)}\big)^k.
\end{gather*}
It follows that the vectors in~\eqref{SVformula1} are the eigenvectors of $D$:
\begin{gather*}
D \big({\cal S}^{(2n)}\big)^k \ket{\delta,p} = (\delta + 2nk) \big({\cal S}^{(2n)}\big)^k \ket{\delta,p},
\\
D \big({\cal S}^{(2n+1)}\big)^k \ket{\delta,p} = (\delta + (2n+1)k) \big({\cal S}^{(2n+1)}\big)^k \ket{\delta,p}.
\end{gather*}
Finally, from the commutation relations
\begin{gather*}
\big[C, {\cal S}^{(2n)}\big]  =  (\ell+2n) P_{\ell-2n+1} (p - P_{\ell} ),
\\ \big[C, {\cal S}^{(2n+1)}\big]  =  \left\{p(\ell +2n+1) P_{\ell-2n} - \frac{(\ell+2n)(\ell+2n+1)}{(n+\frac{1}{2})
(\ell+1)} P_{\ell-1} P_{\ell-2n+1} \right.
\\
\hphantom{\big[C, {\cal S}^{(2n+1)}\big]  = }{}
  +  \frac{(-1)^n}{2n+1}  \frac{\ell! (\ell+2n+1)!}{((\ell+n)!)^2}  P_{\ell-n}^2
\\
\left. \hphantom{\big[C, {\cal S}^{(2n+1)}\big]  = }{} +
\sum\limits_{j=1}^{n-1} \frac{(-1)^j}{n+\frac{1}{2}}  \frac{\ell!  (\ell+2n+1)!}{(\ell+j)!  (\ell+2n-j)!}
P_{\ell-2n+j}  P_{\ell-j} \right\} (p - P_{\ell} ),
\end{gather*}
one can immediately see that $ C ({\cal S}^{(2n)})^k \ket{0} = C ({\cal S}^{(2n+1)})^k \ket{0}=0$.
\end{proof}

\begin{proposition}
\label{prop4}
Following are the singular vectors in $V^{\delta,p}_{\ell}$ for $ \ell \geq 2$
\begin{gather}
\big({\cal T}^{(2n+1)}\big)^k \ket{\delta,p},
\qquad
k = 1, 2, \dots,
\label{SVformula4}
\end{gather}
where $ n $ takes a~value of positive integer.
The maximal value of~$n$ is determined by $ \ell $ and~$N$ in such a~way that $ {\cal T}^{(2n+1)} $ given below is
well-defined
\begin{gather}
    {\cal T}^{(2n+1)} = P_{\ell-1}  \big((\ell+2)  P_{\ell-1}^2 - 2p  (\ell+1)  P_{\ell-2}\big)^n
\nonumber
\\
\hphantom{{\cal T}^{(2n+1)} =}{}
+ \sum\limits_{j=0}^{n-1} c_j  P_{\ell-1}^{2(n-j-1)}  P_{\ell-2}^j  P_{\ell-3} + \sum\limits_{j=1}^n d_j
P_{\ell-1}^{2n-2j+1}  P_{\ell-2}^j,
\label{SVformula5}
\end{gather}
where
\begin{gather*}
c_j = \binom{n-1}{j}  \left(- \frac{2p(\ell+1)}{\ell+2} \right)^j  c_0,
\qquad
d_1 = -p(\ell+1)  (\ell+2)^{n-1}.
\end{gather*}
Other coefficients $ c_0$,
$d_j$ $(j \leq 1 \leq n) $ are determined by the relations
\begin{gather*}
    (j+1)  (\ell+2)  d_{j+1} + (2n-2j+1)  p  (\ell+1) d_j + (\ell+3)  c_{j-1}
\nonumber
\\
\qquad{}
+ \binom{n}{j} (-2)^j  (p(\ell+1))^{j+1}  (\ell+2)^{n-j} = 0,
\qquad
j = 1, 2, \dots, n
%\label{SVformula6}
\end{gather*}
with $d_{n+1} = 0$.
\end{proposition}

\begin{proof}
One can prove this in a~manner similar to Proposition~\ref{prop3}.
Due to the commutativity of $P_{n} $ we can immediately see that the vectors in~\eqref{SVformula4} are annihilated by $
P_a \in {\mathfrak g}_{\ell}^- $ and are eigenvectors of $ P_{\ell} $ with the eigenvalue~$p$.
Using the def\/ining commutation relations~\eqref{nonvanishing-commutators} the following relations can be verif\/ied
without dif\/f\/iculty:
\begin{gather}
    D \big({\cal T}^{(2n+1)}\big)^k \ket{\delta,p} = (\delta + (2n+1)k) \big({\cal T}^{(2n+1)}\big)^k  \ket{\delta,p},
%\label{T-formula1}
\\
    \big[C, {\cal T}^{(2n+1)}\big] = F(P_{\ell-1},P_{\ell-2},P_{\ell-3})  (P_{\ell}-p),
\label{T-formula2}
\end{gather}
where $ F(x,y,z) $ is a~function of three variables and we do not need the explicit form of it.
It follows from~\eqref{T-formula2} that $ C \big({\cal T}^{(2n+1)}\big)^k \ket{\delta,p} = 0$.
Thus the vectors in~\eqref{SVformula4} satisfy the def\/inition of singular vector.
\end{proof}

\subsection[$p=0$]{$\boldsymbol{p=0}$}

In this case singular vectors in $ V^{\delta,0}_{\ell} $ may exist for all values of $\ell $ and $ N$.
We have the singular vectors inherited from the case of $ p \neq 0$.
\begin{proposition}
\label{prop5}
Following are the singular vectors in $ V^{\delta,0}_{\ell}$
\begin{gather}
P_{\ell-1}^k \ket{\delta,0},
\qquad
\big(\tilde{\cal S}^{(2n)}\big)^k \ket{\delta,0},
\qquad
\big(P_{\ell-1} \tilde{\cal S}^{(2n)}\big)^k \ket{\delta,0},
\qquad
k = 1, 2, \dots,
\label{SVp0-fom1}
\end{gather}
where $ n $ takes the value of a~positive integer.
The maximal value of~$n$ is determined by $ \ell $ and~$N$ in such a~way that $ \tilde{\cal S}^{(2n)} $ given below is
well-defined
\begin{gather*}
\tilde{\cal S}^{(2n)} = \sum\limits_{j=1}^{n-1} \frac{(-1)^{n+j}  2  ((\ell+n)!)^2}{(\ell+j)!  (\ell+2n-j)!}
P_{\ell-2n+j}  P_{\ell-j} + P_{\ell-n}^2.
%\label{SVp0-fom2}
\end{gather*}
\end{proposition}

\begin{proof}
It is easy to verify the relations:
\begin{gather*}
\big[D, P_{\ell-1}^k\big] = k  P_{\ell-1}^k,
\qquad
\big[C, P_{\ell-1}^k\big] = k  (\ell+1)  P_{\ell-1}^{\ell-1}  P_{\ell}.
\end{gather*}
It follows immediately that $ P_{\ell-1}^k \ket{\delta,0} $ is a~singular vector with $ \delta + k $ as the eigenvalue
of~$D$.
Next we set $ p = 0 $ at~\eqref{SVformula2},~\eqref{SVformula3} and~\eqref{SVformula5}.
Then we f\/ind the following reduction (up to overall constant):
\begin{gather*}
{\cal S}^{(2n)}  \to \tilde{\cal S}^{(2n)},
\qquad
{\cal S}^{(2n+1)} \to P_{\ell-1} \tilde{\cal S}^{(2n)},
\qquad
{\cal T}^{(2n+1)} \to P_{\ell-1}^{2n+1}.
\end{gather*}
This means that the vectors in~\eqref{SVp0-fom1} are singular vectors.
It is also an easy task to verify directly that the vectors $ (\tilde{\cal S}^{(2n)})^k \ket{\delta,0}$, $(P_{\ell-1}
\tilde{\cal S}^{(2n)})^k \ket{\delta,0} $ satisfy the def\/inition of singular vec\-tor~\eqref{SV-def}.
\end{proof}

\section[Differential equations symmetric under the kinematical transformations generated by ${\mathfrak g}_{\ell}$]{Dif\/ferential equations symmetric\\ under the kinematical transformations generated by $\boldsymbol{{\mathfrak g}_{\ell}}$}
\label{SED:EQS}

\subsection{General formalism}
\label{GeneSche}

The singular vectors obtained in previous section can be used to derive partial dif\/ferential equations having particular
symmetries.
The symmetries are generated by $ {\mathfrak g}_{\ell}$, i.e., the symmetry group is the exponentiation of $ {\mathfrak
g}_{\ell}$, and the partial dif\/ferential equations are invariant under the change of independent variables, i.e., the
kinematical symmetries, caused by the group.
This can be done by applying the method developed for real semisimple Lie groups in~\cite{Dobrev}.
In this subsection we give a~brief review of the method with suitable modif\/ication for the present case (see
also~\cite{ADD,ADDS, AiKimSeg}).

The basic idea is to realize the Verma modules in a~space of $C^{\infty}$-class functions.
Let $G$ be a~complex semisimple Lie group and $ \mathfrak g $ its Lie algebra.
The Lie algebra $ \mathfrak g $ has the triangular decomposition $ {\mathfrak g} = {\mathfrak g}^+ \oplus {\mathfrak
g}^0 \oplus {\mathfrak g}^-$.
The corresponding decomposition of $ G $ is denoted by $ G = G_+ G_0 G_-$.
Consider the space of $C^{\infty}$-class functions on~$G$ having the property called \textit{right covariance}:
\begin{gather*}
C_{\Lambda} = \big\{ f \in C^{\infty}(G) \, | \, f(gxg_-) = e^{\Lambda(H)} f(g) \big\},
%\label{RightCovariance}
\end{gather*}
where $ \Lambda \in {\mathfrak g}^* $ (algebra dual to ${\mathfrak g}_{\ell}$), $ g \in G$, $H \in {\mathfrak g}^0$, $x
= e^H \in G_0$, $g_-\in G_-$.
Because of the right covariance, the functions of $ C_{\Lambda} $ are actually function on $ G/B $ with $ B = G_0 G_-$,
or on $ G_+$.
We keep using the same notations for the restricted representation space of functions on~$ G_+$.
Then one may def\/ine a~representation $T^{\Lambda}$ of $ G $ by a~left regular action on~$ C_{\Lambda}$:
\begin{gather*}
\big(T^{\Lambda}(g) f\big)(g') = f\big(g^{-1}g'\big),
\qquad
g, g' \in G.
%\label{LeftRegRep}
\end{gather*}
The inf\/initesimal generator of this action, which is the standard left action of~$ \mathfrak g $ on~$ C_{\Lambda}$,
gives a~vector f\/ield representation of~$ \mathfrak g$ on~$ C_{\Lambda}$:
\begin{gather*}
\pi_L(X) f(g) = \left.
\frac{d}{d\tau} f\big(e^{-\tau X} g\big) \right|_{\tau=0},
\qquad
X \in {\mathfrak g}, \quad g \in G.
%\label{LeftAction-def}
\end{gather*}

We introduce the right action of $ \mathfrak g $ on $ C_{\Lambda} $ by the standard formula:
\begin{gather*}
\pi_R(X) f(g) = \left.
\frac{d}{d\tau} f\big(g e^{\tau X}\big) \right|_{\tau = 0},
\qquad
X \in {\mathfrak g}, \quad g \in G.
%\label{RightAction-def}
\end{gather*}
One may show by the right covariance that the function $ f \in C_{\Lambda} $ has the properties of lowest weight vector:
\begin{gather*}
\pi_R(H) f(g) = \Lambda(H) f(g),
\qquad
\pi_R(X) f(g) = 0,
\qquad
H \in {\mathfrak g}^0, \quad X \in {\mathfrak g}^-.
%\label{LWfunction-RightAction}
\end{gather*}
This allows us to realize the Verma module $ V^{\Lambda} \simeq U({\mathfrak g}^+) v_0 $ with the lowest weight vector $
v_0 $ in terms of the function in $ C_{\Lambda} $ and dif\/ferential operators $ \pi_R(X)$, $X \in {\mathfrak g}^+$.

Now suppose that the Verma module $ V^{\Lambda} $ has a~singular vector.
The general structure of a~singular vector is
\begin{gather*}
v_s = {\cal P}(X_1, X_2, \dots, X_s) v_0,
\qquad
X_k \in {\mathfrak g}^+,
%\label{gene-str-SV}
\end{gather*}
where $\cal P$ denotes a~homogeneous polynomial in its variables.
The singular vector~$ v_s $ induces the Verma module $ V^{\Lambda'} \simeq U({\mathfrak g}^+) v_s $ with the lowest
weight~$ \Lambda'$.
Thus the dif\/ferential operator $ \pi_R({\cal P}) $ is an intertwining operator between the two representation spaces $
C_{\Lambda} $ and $ C_{\Lambda'}$, i.e.,
\begin{gather}
\pi_R({\cal P})  T^{\Lambda}(g) = T^{\Lambda'}(g)  \pi_R({\cal P}).
\label{DiffOpInter}
\end{gather}
Suppose that the operator $ \pi_R({\cal P}) $ has a~nontrivial kernel
\begin{gather}
\pi_R({\cal P}) \psi = 0,
\label{InvEq-def}
\end{gather}
for some function $ \psi $ on $ G_+$.
The intertwining property~\eqref{DiffOpInter} assures that the equation~\eqref{InvEq-def} is invariant under the
kinematical transformations by $ G$.
\begin{gather*}
\pi_R({\cal P})  T^{\Lambda}(g)  \psi = T^{\Lambda'}(g)  \pi_R({\cal P})  \psi = 0.
\end{gather*}

\subsection{Hierarchies of dif\/ferential equations}

Now let us apply the scheme in Section~\ref{GeneSche} to the group generated by $ {\mathfrak g}_{\ell}$.
We parametrize an element of $ G_+ $ as $g = \exp(tH)  \exp\Big(\sum\limits_{n=0}^{\ell-1} x_n P_n
\Big)$.
Then the right action of $ {\mathfrak g}_{\ell}^+ $ yields
\begin{gather*}
\pi_R(P_n) = \frac{\partial}{\partial x_n},
\qquad
\pi_R(H) = \frac{\partial}{\partial t} + \sum\limits_{j=1}^{\ell-1} j x_j \frac{\partial}{\partial x_{j-1}}.
%\label{RAction-formula}
\end{gather*}
From~\eqref{InvEq-def}, Propositions~\ref{prop3} and~\ref{prop4} we obtain the following hierarchies of
partial dif\/ferential equations.

\begin{proposition}
\label{prop6}
If $ p \neq 0 $ then the following equations are invariant $($in the sense of Section~{\rm \ref{GeneSche})} under the group generated
by ${\mathfrak g}_{\ell}$
\begin{gather}
    \left(p \frac{\partial}{\partial x_{\ell-2n}} + \sum\limits_{j=1}^{n-1} a_j \frac{\partial^2}{\partial x_{\ell-2n+j}
\partial x_{\ell-j}} + \frac{(-1)^n}{2} \frac{\ell! (\ell+2n)!}{((\ell+n)!)^2} \frac{\partial^2}{\partial x_{\ell-n}^2}
\right)^k \psi(x) = 0,
\label{inveqs1}
\\
    \left(p^2 \frac{\partial}{\partial x_{\ell-2n-1}} - \frac{\ell+2n+1}{(n+\frac{1}{2}) (\ell+1)}
\frac{\partial}{\partial x_{\ell-1}}   \left(\sum\limits_{j=1}^{n-1} a_j \frac{\partial^2}{\partial x_{\ell-2n+j}
\partial x_{\ell-j}} + \frac{(-1)^n}{2} \frac{\ell! (\ell+2n)!}{((\ell+n)!)^2} \frac{\partial^2}{\partial x_{\ell-n}^2}
\right) \right.\!
\nonumber
\\
\left.\qquad
{}
- \frac{\ell+2n+1}{\ell+1} p \frac{\partial^2}{\partial x_{\ell-1} x_{\ell-2n}} - \sum\limits_{j=1}^{n-1} b_j
\frac{\partial^2}{\partial x_{\ell-2n+j} x_{\ell-j-1}} \right)^k \psi(x) = 0,
%\label{inveqs2}
\\
    \left(\frac{\partial}{\partial x_{\ell-1}} \left((\ell+2) \frac{\partial^2}{\partial x_{\ell-1}^2} - 2p (\ell+1)
\frac{\partial}{\partial x_{\ell-2}} \right)^n + \sum\limits_{j=0}^{n-1} c_j \left(\frac{\partial}{\partial x_{\ell-1}} \right)^{2(n-j-1)} \left(\frac{\partial}{\partial
x_{\ell-2}} \right)^j \frac{\partial}{\partial x_{\ell-3}}\right.\!
\nonumber
\\
\left.\qquad
{}
+ \sum\limits_{j=1}^n d_j \left(\frac{\partial}{\partial
x_{\ell-1}} \right)^{2n-2j+1} \left(\frac{\partial}{\partial x_{\ell-2}} \right)^j \right)^k \psi(x) = 0.
\label{inveqs3}
\end{gather}
\end{proposition}

We have obtained highly nontrivial dif\/ferential equations.
To have a~close look at the equations, we give examples of the hierarchies of equations for $ n = 1, 2$.
For $n=1 $, the equations~\eqref{inveqs1}--\eqref{inveqs3} are as follows
\begin{gather*}
    \left(p \frac{\partial}{\partial x_{\ell-2}} - \frac{\ell+2}{2(\ell+1)} \frac{\partial^2}{\partial x_{\ell-1}^2}
\right)^k \psi(x) = 0,
%\label{hierarchyEg1}
\\
    \left(p^2 \frac{\partial}{\partial x_{\ell-3}} - \frac{\ell+3}{\ell+1} p \frac{\partial^2}{\partial x_{\ell-2}
\partial x_{\ell-1}} + \frac{(\ell+2) (\ell+3)}{3(\ell+1)^2} \frac{\partial^3}{\partial x_{\ell-1}^3} \right)^k \psi(x)
= 0,
%\label{hierarchyEg2}
\\
    \left((\ell+2) \frac{\partial^3}{\partial x_{\ell-1}^3} - 3(\ell+1) p \frac{\partial^2}{\partial x_{\ell-1} \partial
x_{\ell-2}} + \frac{3(\ell+1)^2}{\ell+3} p^2 \frac{\partial}{\partial x_{\ell-3}} \right)^k \psi(x) = 0.
%\label{hierarchyEg3}
\end{gather*}
The corresponding equations for $ n = 2 $ are given~by
\begin{gather*}
    \left(p \frac{\partial}{\partial x_{\ell-4}} - \frac{\ell+4}{\ell+1} \frac{\partial^2}{\partial x_{\ell-3} \partial
x_{\ell-1}} + \frac{(\ell+3)(\ell+4)}{2(\ell+1)(\ell+2)} \frac{\partial^2}{\partial x_{\ell-2}^2} \right)^k \psi(x) = 0,
%\label{hierarchyEG4}
\\     \left(p^2 \frac{\partial}{\partial x_{\ell-5}} - \frac{\ell+5}{\ell+1} p \frac{\partial^2}{\partial
x_{\ell-4} \partial x_{\ell-1}} + \frac{(\ell+4)(\ell+5)}{5(\ell+1)(\ell+2)} p \frac{\partial^2}{\partial x_{\ell-3}
\partial x_{\ell-2}} \right.
\nonumber
\\
    \left.
\qquad{}
+ \frac{2(\ell+4)(\ell+5)}{5 (\ell+1)^2} \frac{\partial^3}{\partial x_{\ell-3} \partial x_{\ell-1}^2} -
\frac{(\ell+3)(\ell+4)(\ell+5)}{5(\ell+1)^2(\ell+2)} \frac{\partial^3}{\partial x_{\ell-2} \partial x_{\ell-1}}
\right)^k \psi(x) = 0,
\nonumber
\\
    \left((\ell+2)^2 \frac{\partial^5}{\partial x_{\ell-1}^5} - 5(\ell+1)(\ell+2) p \frac{\partial^4}{\partial
x_{\ell-2} \partial x_{\ell-1}^3} + \frac{3(\ell+1)^2 (\ell+2)}{\ell+3} p^2 \frac{\partial^3}{\partial x_{\ell-1}^2
\partial x_{\ell-3}} \right.
\nonumber
\\
    \left.
\qquad{}
+ 6 (\ell+1)^2 p^2 \frac{\partial^3}{\partial x_{\ell-1} \partial x_{\ell-2}^2} -\frac{6(\ell+1)^3}{\ell+3} p^3
\frac{\partial^2}{\partial x_{\ell-2} \partial x_{\ell-3}} \right)^k \psi(x) = 0.
%\label{hierarchyEG6}
\end{gather*}

By a~similar method we obtain, from~\eqref{InvEq-def} and Proposition~\ref{prop5}, the invariant equations for $ p = 0$.
\begin{proposition}
\label{prop7}
If $ p = 0 $, then the following equations are invariant $($in the sense of Section~{\rm \ref{GeneSche})} under the group generated
by ${\mathfrak g}_{\ell}$
\begin{gather*}
    \left(\frac{\partial}{\partial x_{\ell-1}} \right)^k \psi(x) = 0,
%\label{inveqs4}
\\
    \left(\sum\limits_{j=1}^{n-1} \frac{(-1)^{n+j} 2 ((\ell+n)!)^2}{(\ell+j)! (\ell+2n-j)!} \frac{\partial^2}{\partial
x_{\ell-2n+j} \partial x_{\ell-j}} + \frac{\partial^2}{\partial x_{\ell-n}^2} \right)^k \psi(x) = 0,
%\label{inveqs5}
\\
    \left(\frac{\partial}{\partial x_{\ell-1}} \right)^k \left(\sum\limits_{j=1}^{n-1} \frac{(-1)^{n+j} 2
((\ell+n)!)^2}{(\ell+j)! (\ell+2n-j)!} \frac{\partial^2}{\partial x_{\ell-2n+j} \partial x_{\ell-j}} +
\frac{\partial^2}{\partial x_{\ell-n}^2} \right)^k \psi(x) = 0.
%\label{inveqs6}
\end{gather*}
\end{proposition}

\section[Irreducible lowest weight modules of ${\mathfrak g}_{\ell}$]{Irreducible lowest weight modules of $\boldsymbol{{\mathfrak g}_{\ell}}$}
\label{SEC:IRREP}

We have shown that the Verma modules over $ {\mathfrak g}_{\ell} $ are reducible in many cases
(Proposition~\ref{prop2}).
It is known that the Verma module is, in a~sense, the largest lowest weight module.
That is, one can derive all irreducible lowest weight modules starting from the Verma module $ V^{\delta,p}_{\ell}$.
The purpose of this section is to obtain some types of irreducible lowest weight modules explicitly.
Our results are summarized in the next theorem.

\begin{theorem}
\label{theorem2}
The lowest weight modules over ${\mathfrak g}_{\ell}$ given below are irreducible:
\begin{itemize}\itemsep=0pt
\item $ p \neq 0 $
\begin{enumerate}\itemsep=0pt
\item[$i)$] the Verma module $ V^{\delta,p}_1 $ for $ \ell = 1$;
\item[$ii)$] the quotient module $ V_{\ell}^{(\ell)} $ for $ \ell \geq 2$.
This module is infinite-dimensional with the basis vectors $ H^k P_{\ell-1}^m \big|u_0^{(\ell)}\big\rangle $, where $ k$, $m $ are
nonnegative integers.
See Lemma~{\rm \ref{lemma5}} for the definition of the lowest weight vector $ \big| u_0^{(\ell)}\big\rangle$.
\end{enumerate}
\item $ p = 0 $
\begin{enumerate}\itemsep=0pt
\item[$i)$] if $ 2\delta +k = 0 $ for a~nonnegative integer~$k$ then the module isomorphic to the $ k+1$ $(= 2|\delta|+1)$
dimensional module of $ {\mathfrak{sl}}(2,{\mathbb R})$;
\item[$ii)$] if $ 2\delta +k \neq 0 $ for any nonnegative integer~$k$ then the module isomorphic to the infinite-dimensional
module of $ {\mathfrak{sl}}(2,{\mathbb R})$.
\end{enumerate}
\end{itemize}
\end{theorem}

Theorem~\ref{theorem2} coincides with the results in~\cite{LuMazoZhao}.

\subsection[Proof for $p \neq 0$]{Proof for $\boldsymbol{p \neq 0}$}
\label{PFqn0}

The Verma modules $ V^{\delta,p}_{\ell} $ are reducible for $ \ell \geq 2 $ so we restrict ourselves to $ \ell \geq 2$.
We consider the quotient module $ V^{\delta,p}_{\ell}/{\cal I}^{(2)} $ where $ {\cal I}^{(2)} $ is the largest
${\mathfrak g}_{\ell}$-submodule of $ V^{\delta,p}_{\ell}$.
Since there is no singular vectors in the $ N = 1 $ subspace $ (V^{\delta,p}_{\ell})_1$, $ {\cal I}^{(2)} $ will be
induced by the singular vector in the $ N = 2 $ subspace.
\begin{lemma}
\label{lemma3}
There exists precisely one $($up to an overall constant$)$ singular vector in the $ N = 2 $ subspace $
(V^{\delta,p}_{\ell})_2$.
This singular vector is given~by
\begin{gather*}
\big|v_s^{(2)}\big\rangle = (2p (\ell+1) P_{\ell-2} - (\ell+2) P_{\ell-1}^2) \ket{\delta,p}.
%\label{SVN2}
\end{gather*}
\end{lemma}

\begin{proof}
The basis of $ \big(V^{\delta,p}_{\ell}\big)_2 $ is $ \ket{2,\underaccent{\tilde}{0}}, \ket{1,\underaccent{\tilde}{\epsilon}_1},
\ket{0,2\underaccent{\tilde}{\epsilon}_1}, \ket{0,\underaccent{\tilde}{\epsilon}_2}$.
The singular vector $ \big|v_s^{(2)}\big\rangle$ is a~linear combination of the basis:
\begin{gather*}
\big|v_s^{(2)}\big\rangle  = c_1 \ket{2,\underaccent{\tilde}{0}} + c_2 \ket{1,\underaccent{\tilde}{\epsilon}_1} + c_3
\ket{0,2\underaccent{\tilde}{\epsilon}_1} + c_4 \ket{0,\underaccent{\tilde}{\epsilon}_2}.
\end{gather*}
It must satisfy the condition
\begin{gather*}
0 = P_{\ell+1} \big|v_s^{(2)}\big\rangle  = 2 (\ell+1)  p   c_1 \ket{1,\underaccent{\tilde}{0}} + (\ell+1)   (\ell  c_1 + p
c_2) \ket{0,\underaccent{\tilde}{\epsilon}_1}.
\end{gather*}
Thus $ c_1 = c_2 = 0$.
Furthermore, the condition $ C \big|v_s^{(2)}\big\rangle  = 0 $ yields the relation
\begin{gather*}
2 (\ell+1)   p   c_3 + (\ell+2) c_4 = 0.
\end{gather*}
This proves Lemma~\ref{lemma3}.
\end{proof}

Def\/ine $ {\cal I}^{(2)} = U({\mathfrak g}_{\ell}^{+}) \big|v_s^{(2)}\big\rangle $, then $ {\cal I}^{(2)} $ is the largest $
{\mathfrak g}_{\ell}$-submodule in $ V^{\delta,p}_{\ell}$.
Let $ \big|u_0^{(2)}\big\rangle  $ be the lowest weight vector in $ V^{\delta,p}_{\ell}/{\cal I}^{(2)}$.
Then
\begin{gather*}
    D \big|u_0^{(2)}\big\rangle  = \delta \big|u_0^{(2)}\big\rangle ,
\qquad
P_{\ell} \big|u_0^{(2)}\big\rangle  = p \big|u_0^{(2)}\big\rangle ,
\nonumber
\\
    X \big|u_0^{(2)}\big\rangle  = 0, \qquad X \in {\mathfrak g}_{\ell}^-,
\qquad
P_{\ell-2} \big|u_0^{(2)}\big\rangle  = \frac{\ell+2}{2p (\ell+1)} P_{\ell-1}^2 \big|u_0^{(2)}\big\rangle .
%\label{LWVQuo2}
\end{gather*}
It follows that the basis of $ V^{\delta,p}_{\ell}/{\cal I}^{(2)} $ is given~by
\begin{gather*}
\big|k,\underaccent{\tilde}{m}^{(2)}\big\rangle  = H^k   P_{\ell-1}^{m_1}  P_{\ell-3}^{m_3} \cdots P_{0}^{m_{\ell}}
\big|u_0^{(2)}\big\rangle ,
%\label{BasisQuo2}
\end{gather*}
where $ \underaccent{\tilde}{m}^{(2)} = (m_1, 0, m_3, \dots,m_{\ell}) \in {\mathbb R}^{\ell}$.
Observing the relation
\begin{gather*}
D \big|k,\underaccent{\tilde}{m}^{(2)}\big\rangle  = \left(\delta+k +m_1 + \sum\limits_{i=3}^{\ell}i m_i\right)
\big|k,\underaccent{\tilde}{m}^{(2)}\big\rangle ,
\end{gather*}
we def\/ine the level $ N^{(2)} $ in the quotient space $ V^{\delta,p}_{\ell}/{\cal I}^{(2)} $ by $  N^{(2)}
= k + m_1 + \sum\limits_{i=3}^{\ell} i m_{i}$.
The vectors $ \{
\ket{1,\underaccent{\tilde}{0}},
\ket{0,\underaccent{\tilde}{\epsilon}_1}
\} $ and $ \{
\ket{2,\underaccent{\tilde}{0}},
\ket{1,\underaccent{\tilde}{\epsilon}_1},
\ket{0,2\underaccent{\tilde}{\epsilon}_1}
\}$ form a~basis of $ N^{(2)} = 1 $ and $ N^{(2)} = 2 $ subspaces of $ V^{\delta,p}_{\ell}/{\cal I}^{(2)}$,
respectively.
The Shapovalov determinant for $ N^{(2)} = 1 $ subspace is same as~\eqref{SEC:VM} and given by $ (\ell+1)^2 p^2$.
While that for $ N^{(2)} = 2 $ subspace is calculated as follows
\begin{gather*}
    \left|
\begin{matrix} \bracket{2,\underaccent{\tilde}{0}}{2,\underaccent{\tilde}{0}} &
\bracket{2,\underaccent{\tilde}{0}}{1,\underaccent{\tilde}{\epsilon}_1} &
\bracket{2,\underaccent{\tilde}{0}}{0,2\underaccent{\tilde}{\epsilon}_1}
\vspace{1mm}\\ \bracket{1,\underaccent{\tilde}{\epsilon}_1}{2,\underaccent{\tilde}{0}} &
\bracket{1,\underaccent{\tilde}{\epsilon}_1}{1,\underaccent{\tilde}{\epsilon}_1} &
\bracket{1,\underaccent{\tilde}{\epsilon}_1}{}
\vspace{1mm}\\ \bracket{0,2\underaccent{\tilde}{\epsilon}_1}{2,\underaccent{\tilde}{0}} &
\bracket{0,2\underaccent{\tilde}{\epsilon}_1}{1,\underaccent{\tilde}{\epsilon}_1} &
\bracket{0,2\underaccent{\tilde}{\epsilon}_1}{0,2\underaccent{\tilde}{\epsilon}_1}
\end{matrix}
\right|
\\
\qquad{}
= \left|
\begin{matrix} 4(2\delta+1) \delta & 2(2\delta+1)(\ell+1)p & 2(\ell+1)^2 p^2
\vspace{1mm}\\ 2(2\delta+1)(\ell+1)p & (\ell+1)^2 p^2 & 0
\vspace{1mm}\\ 2(\ell+1)^2 p^2 & 0 & 0
\end{matrix}
\right| = -4(\ell+1)^6 p^6.
\end{gather*}
Therefore there exist no singular vectors in $ N^{(2)} = 1, 2 $ subspaces.
On the other hand, one f\/inds a~singular vector in the $ N^{(2)} = 3 $ subspace.

\begin{lemma}
\label{lemma4}
There exists precisely one $($up to overall constant$)$ singular vector in the $ N^{(2)} = 3 $ subspace of $
V^{\delta,p}_{\ell}/{\cal I}^{(2)}$.
The singular vector is given~by
\begin{gather*}
\big|v_s^{(3)}\big\rangle  = \big(3! p^2 (\ell+1)^2  P_{\ell-3} - (\ell+2)  (\ell+3)  P_{\ell-1}^3\big) \big|u_0^{(2)}\big\rangle .
%\label{SVN3}
\end{gather*}
\end{lemma}

\begin{proof}
The lemma can be proved in a~way exactly similar to Lemma~\ref{lemma3}.
The basis of the level $ N^{(2)} = 3 $ subspace is given~by
\begin{gather*}
\ket{3,\underaccent{\tilde}{0}},
\qquad
\ket{2,\underaccent{\tilde}{\epsilon}_1},
\qquad
\ket{1,2\underaccent{\tilde}{\epsilon}_1},
\qquad
\ket{0,3\underaccent{\tilde}{\epsilon}_1},
\qquad
\ket{0,\underaccent{\tilde}{\epsilon}_3}.
\end{gather*}
The singular vector $ \big|v_s^{(3)}\big\rangle  $ is a~linear combination of the basis:
\begin{gather*}
\big|v_s^{(3)}\big\rangle  = c_1 \ket{3,\underaccent{\tilde}{0}} + c_2 \ket{2,\underaccent{\tilde}{\epsilon}_1} + c_3
\ket{1,2\underaccent{\tilde}{\epsilon}_1} + c_4 \ket{0,3\underaccent{\tilde}{\epsilon}_1} + c_5
\ket{0,\underaccent{\tilde}{\epsilon}_3}.
\end{gather*}
It must satisfy the condition
\begin{gather*}
0  =  P_{\ell+1} \big|v_s^{(3)}\big\rangle
\\
\hphantom{0}{}
 =  3 p c_1 \ket{2,\underaccent{\tilde}{0}} + (3\ell c_1 + 2p c_2) \ket{1,\underaccent{\tilde}{\epsilon}_1} +
\left(\frac{(\ell+2) \ell (\ell-1)}{2(\ell+1)p} c_1 + \ell c_2 + (\ell+1) p c_3 \right) \ket{0,
2\underaccent{\tilde}{\epsilon}_1}.
\end{gather*}
Thus $ c_1 = c_2 = c_3 = 0$.
Furthermore, the condition $ C \big|v_s^{(3)}\big\rangle  = 0 $ yields the relation
\begin{gather*}
3 (\ell+1) p c_4 + \frac{(\ell+3) (\ell+2)}{2(\ell+1)p} c_5 = 0.
\end{gather*}
This proves  Lemma~\ref{lemma4}.
\end{proof}

The vector $ \big|v_s^{(3)}\big\rangle  $ is singular only in the quotient space $ V^{\delta,p}_{\ell}/{\cal I}^{(2)} $ and not in
$ V^{\delta,p}_{\ell} $ itself.
Such vector is called \textit{subsingular}~\cite{Dob95,Dob97}.

It follows from Lemma~\ref{lemma4} that the subspace $ {\cal I}^{(3)} = U({\mathfrak g}_{\ell}^+) \big|v_s^{(3)}\big\rangle  $ is
the largest ${\mathfrak g}_{\ell}$-submodule in $ V^{\delta,p}_{\ell}/{\cal I}^{(2)}$.
Now we consider the quotient space $ V^{\delta,p}_{\ell}/{\cal I}^{(2)}/{\cal I}^{(3)}:= \big(V^{\delta,p}_{\ell}/{\cal
I}^{(2)}\big)/{\cal I}^{(3)}$.
The lowest weight vector $ \big|u_0^{(3)}\big\rangle  $ of this quotient space is def\/ined~by
\begin{gather*}
    D \big|u_0^{(3)}\big\rangle  = \delta \big|u_0^{(3)}\big\rangle ,
\qquad
P_{\ell} \big|u_0^{(3)}\big\rangle  = p \big|u_0^{(3)}\big\rangle ,
\qquad
    X \big|u_0^{(3)}\big\rangle  = 0, \qquad X \in {\mathfrak g}_{\ell}^-,
\nonumber
\\
    P_{\ell-2} \big|u_0^{(3)}\big\rangle  = \frac{\ell+2}{2p  (\ell+1)} P_{\ell-1}^2 \big|u_0^{(3)}\big\rangle ,
\qquad
P_{\ell-3} \big|u_0^{(3)}\big\rangle  = \frac{(\ell+2) (\ell+3)}{3!  p^2  (\ell+1)^2} P_{\ell-1}^3\big|u_0^{(3)}\big\rangle .
%\label{LWVQuo3}
\end{gather*}
The basis of $ V^{\delta,p}_{\ell}/{\cal I}^{(2)}/{\cal I}^{(3)} $ is given~by
\begin{gather*}
H^k  P_{\ell-1}^{m_1} P_{\ell-4}^{m_4} \cdots P_0^{m_{\ell}} \big|u_0^{(3)}\big\rangle ,
\end{gather*}
and we def\/ine the level $  N^{(3)} = k + m_1 + \sum\limits_{i=4}^{\ell} i  m_i$.
With this setting one can show the followings:
\begin{enumerate}\itemsep=0pt
\item[i)] the Shapovalov determinant does not vanish for $ 1 \leq N^{(3)} \leq 3 $ so that there exist no singular vectors
in the subspaces with $ N^{(3)} = 1, 2, 3$;
\item[ii)] there exists a~unique singular vector in the level $ N^{(3)} = 4 $ subspace.
\end{enumerate}
By the singular vector in $ N^{(3)} = 4 $ subspace, one can def\/ine the largest ${\mathfrak g}_{\ell}$-submodule $ {\cal
I}^{(4)} $ in $ V^{\delta,p}_{\ell}/{\cal I}^{(2)}/{\cal I}^{(3)} $ and consider the quotient space $
V^{\delta,p}_{\ell}/{\cal I}^{(2)}/{\cal I}^{(3)}/{\cal I}^{(4)}$.

In fact, one can repeat this process until we arrive at $ V^{\delta,p}_{\ell}/{\cal I}^{(2)}/\cdots/{\cal I}^{(\ell)}$.
This is assured by the next lemma:
\begin{lemma}
\label{lemma5}
Suppose that we have arrived at the quotient space $ V_{\ell}^{(\lambda)}:= V^{\delta,p}_{\ell}/{\cal
I}^{(2)}/\cdots/{\cal I}^{(\lambda)}$, $2 \leq \lambda \leq \ell$.
Namely, we have the quotient space with the basis
\begin{gather*}
H^k P_{\ell-1}^{m_1} P_{\ell-\lambda-1}^{m_{\lambda+1}} \cdots P_0^{m_{\ell}} \big|u_0^{(\lambda)}\big\rangle ,
%\label{LWVQuoL}
\end{gather*}
where $ \big|u_0^{(\lambda)}\big\rangle  $ is the lowest weight vector in $ V_{\ell}^{(\lambda)} $, defined~by
\begin{gather}
    D \big|u_0^{(\lambda)}\big\rangle  = \delta \big|u_0^{(\lambda)}\big\rangle ,
\qquad
P_{\ell} \big|u_0^{(\lambda)}\big\rangle  = p \big|u_0^{(\lambda)}\big\rangle ,
\qquad
X \big|u_0^{(\lambda)}\big\rangle  = 0, \qquad X \in {\mathfrak g}_{\ell}^-,
\nonumber
\\      P_{\ell-a} \big|u_0^{(\lambda)}\big\rangle  = \frac{(\ell+2) (\ell+3) \cdots (\ell+a)}{a!   p^{a-1}   (\ell+1)^{a-1}}
P_{\ell-1}^a \big|u_0^{(\lambda)}\big\rangle ,
\qquad
a = 2, 3, \dots, \lambda.
\label{Assumption-lambda}
\end{gather}
Then
\begin{enumerate}\itemsep=0pt
\item[$i)$] $ V_{\ell}^{(\lambda)} $ is the graded vector space:
\begin{gather*}
    V_{\ell}^{(\lambda)} = \bigoplus_{N^{(\lambda)}=0}^{\infty} (V_{\ell}^{(\lambda)})_{N^{(\lambda)}},
\qquad
\big(V_{\ell}^{(\lambda)}\big)_{N^{(\lambda)}} = \big\{ \ket{v} \in V_{\ell}^{(\lambda)} \, | \, D \ket{v} = (\delta +
N^{(\lambda)}) \ket{v} \big\},
\nonumber
\\
    N^{(\lambda)} = k + m_1 + \sum\limits_{i=\lambda+1}^{\ell} i m_i.
%\label{gradedquotientspace}
\end{gather*}
\item[$ii)$] The subspace $ \big(V_{\ell}^{(\lambda)}\big)_{N^{(\lambda)}} $ has a~nonvanishing Shapovalov determinant if $ 1 \leq
N^{(\lambda)} \leq \lambda$.
The Shapovalov determinant is given by $($up to a~sign factor$)$
\begin{gather}
\Delta_{N^{(\lambda)}} = (p  (\ell+1))^{N^{(\lambda)}(N^{(\lambda)}+1)} \prod\limits_{k=0}^{N^{(\lambda)}} k!
\big(N^{(\lambda)}-k\big)!.
\label{KDinQuotient}
\end{gather}
This implies that there exists no singular vectors in the level $ N^{(\lambda)} $ subspaces if $ 1 {\leq} N^{(\lambda)}
{\leq} \lambda$.
\item[$iii)$] If $ 2 \leq \lambda \leq \ell-1$, then there exists precisely one $($up to overall constant$)$ singular vector in the
$ N^{(\lambda)} = \lambda + 1 $ subspace of $ V^{(\lambda)}_{\ell}$.
The singular vector is given~by
\begin{gather}
\big|v_s^{(\lambda{+}1)}\big\rangle  = \big(  (\lambda+1)!  p^{\lambda} (\ell+1)^{\lambda} P_{\ell{-}\lambda{-}1} \!- (\ell+2) (\ell+3)
\cdots (\ell+\lambda+1) P_{\ell{-}1}^{\lambda{+}1}  \big)\! \big|u_0^{(\lambda)}\big\rangle .\!\!\!\!
\label{SVinQuotientSpace}
\end{gather}
\end{enumerate}
\end{lemma}

\begin{proof}
(i) Can be trivially proved.
 (ii) Suppose that $ 1 \leq N^{(\lambda)} \leq \lambda$, then the basis of $
\big(V_{\ell}^{(\lambda)}\big)_{N^{(\lambda)}} $ is given by $ \ket{k} = H^{N^{(\lambda)}-k} P_{\ell-1}^k \big|u_0^{(\lambda)}\big\rangle
$ with $ k = 0, 1, \dots, N^{(\lambda)}$.
The equality in the next equation is up to a~sign factor:
\begin{gather}
\Delta_{N^{(\lambda)}} = \left|
\begin{matrix} \bracket{0}{N^{(\lambda)}} & \bracket{0}{N^{(\lambda)}-1} & \dots & \bracket{0}{0}
\\ \bracket{1}{N^{(\lambda)}} & \bracket{1}{N^{(\lambda)}-1} & \dots & \bracket{1}{0}
\\
\vdots & \vdots & & \vdots
\\
\bracket{N^{(\lambda)}}{N^{(\lambda)}} & \bracket{N^{(\lambda)}}{N^{(\lambda)}-1} & \vdots & \bracket{N^{(\lambda)}}{0}
\end{matrix}
\right|.
\label{KDQuoSpa}
\end{gather}
The lower triangular entry of~\eqref{KDQuoSpa} is given~by
\begin{gather}
\big\langle m \big|  N^{(\lambda)}-k\big\rangle  = \big\langle u_0^{(\lambda)}\big| P_{\ell+1}^m C^{N^{(\lambda)}-m} H^k P_{\ell-1}^{N^{(\lambda)}-k}
\big|u_0^{(\lambda)}\big\rangle ,
\qquad
k < m.
\label{Prod-m-Nk}
\end{gather}
Using the commutation relations
\begin{gather*}
\big[C, H^k\big] = 2k H^{k-1} D + k(k-1) H^{k-1},
\qquad
\big[C, P_{\ell-1}^k\big] = k (\ell+1) P_{\ell-1}^{k-1} P_{\ell},
\end{gather*}
we see that~\eqref{Prod-m-Nk} is a~linear combination of
\begin{gather}
\big\langle u_0^{(\lambda)}\big| P_{\ell+1}^m H^j P_{\ell-1}^{m-j} \big|u_0^{(\lambda)}\big\rangle ,
\qquad
\max\{k-N+m, 0\} \leq j \leq k.
\label{Prod-m-Nk-2}
\end{gather}
The lower bound for~$j$ corresponds to the two possibilities $ N-m \leq k $ or $ N-m > k$.
By the relation
\begin{gather*}
\big[H, P_{\ell+1}^m\big] = -m(\ell+1) P_{\ell+1}^{m-1} P_{\ell},
\end{gather*}
the equation~\eqref{Prod-m-Nk-2} is further reduced to
\begin{gather*}
\big\langle u_0^{(\lambda)}\big| P_{\ell+1}^{m-j} P_{\ell-1}^{m-j} \big|u_0^{(\lambda)}\big\rangle  = 0.
\end{gather*}
The last equality stems from $ m -j > 0 $ which is due to the range of~$j$ and $ k < m$.
Thus all the lower triangular entries of~\eqref{KDQuoSpa} vanish.

The diagonal entries of~\eqref{KDQuoSpa} correspond to~\eqref{Prod-m-Nk} with $ m = k$.
Thus they are expanded into a~linear combination of~\eqref{Prod-m-Nk-2}.
Because of $ m = k $ there exist precisely one term in the expansion which gives the nonvanishing contribution.
Writing the coef\/f\/icient explicitly, that term is given~by
\begin{gather*}
(N^{(\lambda)}-k) ((\ell+1) p)^{N^{(\lambda)}-k} \big\langle u_0^{(\lambda)}\big| P_{\ell+1}^k H^k \big|u_0^{(\lambda)}\big\rangle  = k!
(N^{(\lambda)}-k)!   (p  (\ell+1))^{N^{(\lambda)}}.
\end{gather*}
The formula~\eqref{KDinQuotient} immediately follows from these results.

 (iii) The basis vectors of level $ N^{(\lambda)} = \lambda + 1 $ subspace are $ P_{\ell-\lambda-1}
\big|u_0^{(\lambda)}\big\rangle  $ and $ \ket{k} =$ \linebreak  $H^{\lambda+1-k}   P_{\ell-1}^k \big|u_0^{(\lambda)}\big\rangle$ with $
k=0,1,\dots,\lambda+1$.
The singular vector is a~linear combination of these vectors:
\begin{gather*}
\big|v_s^{(\lambda+1)}\big\rangle  = \sum\limits_{k=0}^{\lambda+1} \alpha_k \ket{k} + \beta P_{\ell-\lambda-1}
\big|u_0^{(\lambda)}\big\rangle .
\end{gather*}
The condition $ P_{\ell+1} \big|v_s^{(\lambda+1)}\big\rangle  = 0 $ yields
\begin{gather*}
    (\lambda+1-n) \left(  p (\ell+1)  \alpha_n + \binom{\ell+1}{2} (\lambda+2-n) \alpha_{n-1} \right) H^{\lambda-n}
P_{\ell-1}^n \big|u_0^{(\lambda)}\big\rangle
\\
\qquad{}
+ \sum\limits_{j=2}^n \binom{\ell+1}{j+1} \frac{(\lambda+1-n+j)!}{(\lambda-n)!}   \alpha_{n-j} H^{\lambda-n} P_{\ell-j}
P_{\ell-1}^{n-j} \big|u_0^{(\lambda)}\big\rangle  = 0,
\qquad
0 \leq n \leq \lambda.
\end{gather*}
It follows that $ \alpha_j = 0 $ for $ 0 \leq j \leq \lambda$.
Thus the singular vector yields
\begin{gather*}
\big|v_s^{(\lambda+1)}\big\rangle  = \big(  \alpha_{\lambda+1} P_{\ell-1}^{\lambda+1} + \beta P_{\ell-\lambda-1}  \big)
\big|u_0^{(\lambda)}\big\rangle .
\end{gather*}
The condition $ C \big|v_s^{(\lambda+1)}\big\rangle  = 0 $ gives the relation
\begin{gather*}
(\lambda+1)!  p^{\lambda} (\ell+1)^{\lambda} \alpha_{\lambda+1} + (\ell+2) (\ell+3) \cdots (\ell+\lambda+1) \beta = 0.
\end{gather*}
This implies the uniqueness of the formula~\eqref{SVinQuotientSpace} of the singular vector.
We remark that if $ \lambda=\ell $ then the vector corresponds to $ P_{\ell-\lambda-1} $ does not exist.
Thus $ \alpha_{\lambda+1} = 0 $, so that no singular vectors at level $ \ell+1$.
\end{proof}

Now we consider the module $ V^{(\ell)}_{\ell} = V^{\delta,p}_{\ell}/{\cal I}^{(2)}/ {\cdots } /{\cal I}^{(\ell)}$.
The basis of this space is $ H^k P_{\ell{-}1}^m \big|u_0^{(\ell)}\big\rangle  $ where $ \big|u_0^{(\ell)}\big\rangle  $ is the lowest weight
vector def\/ined by~\eqref{Assumption-lambda} with $ \lambda = \ell$.
By Lemma~\ref{lemma5}, there exist no singular vectors in the subspaces of $ V^{(\ell)}_{\ell} $ labelled by $
N^{(\ell)} = 1, 2, \dots, \ell$.
For $ N^{(\ell)} \geq \ell+1 $ the singular vectors may be written as
\begin{gather*}
\big|v_s^{(\ell)}\big\rangle  = \sum\limits_{k=0}^{N^{(\ell)}} \alpha_k H^k P_{\ell-1}^{N^{(\ell)}-k} \big|u_0^{(\ell)}\big\rangle .
\end{gather*}

The condition $ P_{\ell+1} \big|v_s^{(\ell)}\big\rangle  = 0 $ yields
\begin{gather*}
\sum\limits_{k=1}^{N^{(\ell)}} \sum\limits_{j=1} \alpha_k \binom{\ell+1}{j} \frac{k!}{(k-j)!} H^{k-j} P_{\ell+1-j}
P_{\ell-1}^{N^{(\ell)}-k} \big|u_0^{(\ell)}\big\rangle  = 0.
\end{gather*}
From the above relation we can see that $ \alpha_k = 0$ $(k \neq 0)$.
We also see that $ \alpha_0 = 0 $ from
\begin{gather*}
0 = C \big|v_s^{(\ell)}\big\rangle  = \alpha_0 N^{(\ell)} (\ell+1) p P_{\ell-1}^{N^{(\ell)}-1} \big|u_0^{(\ell)}\big\rangle .
\end{gather*}
Thus there are no singular vectors in $ V^{(\ell)}_{\ell}$.
That is, $ V^{(\ell)}_{\ell} $ is an irreducible ${\mathfrak g}_{\ell}$-module.

\subsection[Proof for $p=0$]{Proof for $\boldsymbol{p=0}$}

From Proposition~\ref{prop1} if $ p = 0 $, then $ \Delta^{(\ell)}_N = 0 $ for all values of $ \ell $ and $ N$.
Thus there will be singular vectors at level $ N = 1 $ subspace of $ V^{\delta,0}_{\ell} $ and they may be of the form
\begin{gather*}
\ket{v_s} = (\alpha H + \beta P_{\ell-1}) \ket{\delta,0}.
\end{gather*}
It is easy to see that
\begin{gather*}
P_k \ket{v_s} = 0, \qquad k \geq \ell+1,
\qquad
C \ket{v_s} = 2 \alpha \delta \ket{\delta,0}.
\end{gather*}
Thus there are two possibilities:
\begin{enumerate}\itemsep=0pt
\item[a)] $ \alpha = 0 $ and $ \delta $ is arbitrary (so we take $ \delta \neq 0$).
In this case there exists precisely one singular vector at $ N = 1 $ and it is given by $ P_{\ell-1} \ket{\delta,0}$.
\item[b)] $ \delta = 0 $ and $ \alpha $ is arbitrary.
In this case there exists two independent singular vectors given by $ H \ket{0,0}$, and~$P_{\ell-1} \ket{0,0}$.
\end{enumerate}
We treat these cases separately.

(a) $ \delta \neq 0$: The largest ${\mathfrak g}_{\ell}$-submodule in $ V^{\delta,0}_{\ell} $ is $ {\cal
I}^{(1)} = U({\mathfrak g}_{\ell}^+) P_{\ell-1} \ket{\delta,0}$.
The lowest weight vector $ \big|u_0^{(1)}\big\rangle  $ of the quotient space $ V^{\delta,0}_{\ell}/{\cal I}^{(1)} $ is annihilated
by $ P_{\ell-1} $ in addition to any element of~$ {\mathfrak g}_{\ell}^+$.
The basis of $ V^{\delta,0}_{\ell}/{\cal I}^{(1)} $ is $ H^k P_{\ell-2}^{m_2} \cdots P_0^{m_{\ell}} \big|u_0^{(1)}\big\rangle  $ and
we set $  N^{(1)} = k + \sum\limits_{i=2}^{\ell} i   m_i$.
The $ N^{(1)} = 1 $ subspace is spanned by only one vector $ H \big|u_0^{(1)}\big\rangle  $ so it is easy to see that there exists
no singular vectors in $ N^{(1)} = 1 $ subspace.
It is also easy to see that there exists precisely one singular vector in $ N^{(1)} = 2 $ subspace and it is given by $
P_{\ell-2} \big|u_0^{(1)}\big\rangle $.
Thus $ {\cal I}^{(2)} = U({\mathfrak g}_{\ell}^+) P_{\ell-2} \big|u_0^{(1)}\big\rangle  $ is the largest ${\mathfrak
g}_{\ell}$-submodule in $ V^{\delta,0}_{\ell}/{\cal I}^{(1)} $, so that we consider the quotient $
V^{\delta,0}_{\ell}/{\cal I}^{(1)}/{\cal I}^{(2)}$.
Similar to the $ p \neq 0 $ case in Section~\ref{PFqn0} one can repeat this process again and again until $
V^{\delta,0}_{\ell}/{\cal I}^{(1)}/\cdots/{\cal I}^{(\ell)}$.

\begin{lemma}
%\label{lemma6}
Suppose that we arrived at the quotient space $ V_{\ell}^{(\lambda)}:= V^{\delta,0}_{\ell}/{\cal I}^{(1)}/\cdots/{\cal
I}^{(\lambda)}$, $1 \leq \lambda \leq \ell-1$.
Namely, we have the quotient space with the basis
\begin{gather*}
H^k P_{\ell-\lambda-1}^{m_{\lambda+1}} \cdots P_0^{m_{\ell}} \big|u_0^{(\lambda)}\big\rangle ,
%\label{LWVQuoLp0}
\end{gather*}
where $ \big|u_0^{(\lambda)}\big\rangle  $ is the lowest weight vector in $ V_{\ell}^{(\lambda)} $ defined~by
\begin{gather*}
D \big|u_0^{(\lambda)}\big\rangle  = \delta \big|u_0^{(\lambda)}\big\rangle ,
\qquad
C \big|u_0^{(\lambda)}\big\rangle  = P_k \big|u_0^{(\lambda)}\big\rangle  = 0, \qquad \ell-\lambda \leq k \leq 2\ell.
%\label{LWVp0Quotient}
\end{gather*}
Then
\begin{enumerate}\itemsep=0pt
\item[$i)$] $ V_{\ell}^{(\lambda)} $ is the graded vector space:
\begin{gather*}
    V_{\ell}^{(\lambda)} = \bigoplus_{N^{(\lambda)}=0}^{\infty} \big(V_{\ell}^{(\lambda)}\big)_{N^{(\lambda)}},
\qquad
\big(V_{\ell}^{(\lambda)}\big)_{N^{(\lambda)}} = \big\{ \ket{v} \in V_{\ell}^{(\lambda)} \, | \, D \ket{v} = (\delta +
N^{(\lambda)}) \ket{v} \big\},
\nonumber
\\
    N^{(\lambda)} = k + \sum\limits_{i=\lambda+1}^{\ell} i m_i.
%\label{gradedquotientspace2}
\end{gather*}
\item[$ii)$] The subspace $ \big(V_{\ell}^{(\lambda)}\big)_{N^{(\lambda)}} $ does not have any singular vector if, $ 1 \leq
N^{(\lambda)} \leq \lambda$.
\item[$iii)$] There exists precisely one singular vector $ P_{\ell-\lambda-1} \big|u_0^{(\lambda)}\big\rangle  $ at level $ N^{(\lambda)} =
\lambda + 1 $ subspace.
\end{enumerate}
\end{lemma}

\begin{proof}
(i) Can be easily proved.
(ii) The subspace $ (V_{\ell}^{(\lambda)})_{N^{(\lambda)}} $ is spanned by only one vector $ H^{N^{(\lambda)}}
\big|u_0^{(\lambda)}\big\rangle  $ if $ 1 \leq N^{(\lambda)} \leq \lambda$.
It is easy to see that the vector is not annihilated by, for instan\-ce,~$P_{\ell-\lambda}$.

(iii) A~singular vector at level $ N^{(\lambda)} = \lambda + 1 $ subspace is written as
\begin{gather*}
\ket{v_s} = (\alpha H^{\lambda+1} + \beta P_{\ell-\lambda-1}) \big|u_0^{(\lambda)}\big\rangle .
\end{gather*}
From~\eqref{PnHkcomm} if $ 1 \leq \lambda \leq \ell-1 $ we have
\begin{gather*}
P_{\ell} \ket{v_s} \sim \alpha P_{\ell-\lambda-1} \big|u_0^{(\lambda)}\big\rangle ,
\qquad
C \ket{v_s} = \alpha (\lambda+1) (2\delta+\lambda) H^{\lambda} \big|u_0^{(\lambda)}\big\rangle .
\end{gather*}
It follows that $ \alpha = 0 $, and $ \beta $ is arbitrary.
\end{proof}

Now we analyze $ V_{\ell}^{(\ell)} $ in a~more detailed manner.
The basis of $ V_{\ell}^{(\ell)} $ is $ \ket{k}:= H^k \big|u_0^{(\ell)}\big\rangle $.
From~\eqref{PnHkcomm} we see that $ P_n \ket{k} = 0 $ for all~$k$ and~$n$.
Therefore, only the $ {\mathfrak{sl}}(2,{\mathbb R}) $ subalgebra (spanned by $ H$,~$D$,~$C$) of $ {\mathfrak g}_{\ell}
$ has nontrivial action on $ V_{\ell}^{(\ell)}$.
Namely, $ V_{\ell}^{(\ell)} $ is isomorphic to a~${\mathfrak{sl}}(2,{\mathbb R})$-module.
$ V_{\ell}^{(\ell)} $ is also a~graded vector space.
Each subspace (labelled by the positive integer~$k$) is one-dimensional with the basis vector~$\ket{k}$.
Since $ C \ket{k} = k(2\delta+k-1) \ket{k-1}$, if $ 2\delta+k-1 = 0 $ then $ \ket{k} $ is the unique singular vector in
$ V_{\ell}^{(\ell)}$.
The quotient module $ V_{\ell}^{(\ell)}/U({\mathfrak g}_{\ell}^+) \ket{k} $ is $ k$-dimensional irreducible module of $
{\mathfrak{sl}}(2,{\mathbb R)}$.

(b) $ \delta = 0: $ Let $ {\cal J} = U({\mathfrak g}_{\ell}^+) H \ket{0,0} $ and consider the quotient $
V^{0,0}_{\ell}/{\cal J}$.
This space is spanned by $ P_{\ell-1}^{m_1} \cdots P_0^{m_{\ell}} \ket{v_0} $ with the lowest weight vector $ \ket{v_0}$.
Now we repeat the same process as in the previous cases and see that the space shrinks step by step to the linear span
of $ P_{\ell-2}^{m_2} \cdots P_0^{m_{\ell}} \ket{v_0}$, $ P_{\ell-3}^{m_3} \cdots P_0^{m_{\ell}} \ket{v_0}$,  etc.
Finally, we arrive at the irreducible one-dimensional space which gives the trivial representation of $ {\mathfrak
g}_{\ell}$.

\section{Concluding remarks}

The main results of this work are the following: Explicit formulae of singular vectors in $V^{\delta,p}_{\ell}$ over
${\mathfrak g}_{\ell}$ (Propositions~\ref{prop3},~\ref{prop4} and~\ref{prop5}).
Hierarchies of partial dif\/ferential equation symmetric under the transformations generated by $ {\mathfrak g}_{\ell} $
(Propositions~\ref{prop6} and~\ref{prop7}).
Irreducible lowest weight modules of $ {\mathfrak g}_{\ell} $ (Theorem~\ref{theorem2}).
In the present work and in~\cite{AiIsKi,AiKimSeg,LuMazoZhao} the irreducible lowest/highest weight modules of the $ d=1
$ CGA and invariant dif\/ferential equations have been investigated in full detail.
However, the structure of irreducible modules of the CGA for $ d \geq 2 $ (especially higher values of~$\ell$) is still
an open problem.
Another interesting problem may be a~relation of the representation of CGA and orthogonal polynomials.
In this regard we would like to cite~\cite{VineZheda} wherein a~relation between the representation of the Schr\"odinger
\textit{group} and a~discrete matrix orthogonal polynomial has been discussed in detail.
However to the best of our knowledge no such relationship between orthogonal polynomials and the representation of CGA
is known.
A~key observation for such polynomials might be the formulae of singular vectors presented in Section~\ref{Sec:SV}.
Because they are multivariable polynomials on $ P_n$'s and orthogonal with respect to the inner
product~\eqref{innerprod-def}.
A~thorough investigation of this aspect is beyond the scope of the current work and hence will be reported elsewhere.

\vspace{-1.5mm}

\subsection*{Acknowledgements}

\vspace{-1mm}

The work of N.A.~is supported by the grants-in-aid from JSPS (Contract No.26400209).
J.S.~acknowledges the hospitality of OPU, where part of this work was completed.
R.C.~was f\/inancially supported through the MOST grants 102-2811-M-005-025 and 102-2628-M-005-001-MY4 in Taiwan.
He would like to thank Professor Naruhiko Aizawa for the invitation to visit Osaka Prefecture University and also for
the hospitality extended to him during his stay.

\vspace{-3mm}

\pdfbookmark[1]{References}{ref}

\LastPageEnding

\end{document}